\documentclass[a4paper, conference]{IEEEtran}
\IEEEoverridecommandlockouts
\usepackage[nocompress]{cite}
\usepackage{amsmath,amssymb,amsfonts}
\usepackage{algorithmic}
\usepackage{graphicx}
\usepackage{textcomp}
\usepackage{xcolor}
\usepackage{amssymb}  
\usepackage{algorithm}
\usepackage{amsmath}
\usepackage{lipsum}

\begin{document}
 \title{\huge Study of Code-Aided Channel Estimation for Metasurface-Based Holographic MIMO Systems\vspace{-0.05em} }

\author{Roberto C. G. Porto and Rodrigo C. de Lamare \vspace{-0.1em}

\thanks{The authors are with the Centre for Telecommunications Studies, Department of Electrical Engineering, Pontifical Catholic University of Rio de Janeiro. Emails: camara@ime.eb.br, delamare@puc-rio.br}}

\maketitle
 \begin{abstract}
This work proposes an iterative code-aided detection, decoding, and channel estimation scheme for metasurface-based holographic MIMO systems employing stacked intelligent metasurfaces (SIM-RIS) and their fully connected counterparts (BD-SIM-RIS). A novel channel estimation strategy is developed by exploiting low-density parity-check (LDPC) coding in the uplink, enabling both pilot and parity bits of the encoded packet to contribute to the iterative refinement of the channel. In addition, closed-form expressions for the metasurface parameter design are derived and incorporated into an alternating-optimization (AO) procedure. 
Numerical results demonstrate substantial gains in normalized mean square error (NMSE) and bit error rate (BER), with particularly strong improvements observed for BD-SIM-RIS architectures.
\end{abstract}

\begin{IEEEkeywords}
Metasurfaces, SIM-RIS, BD-SIM-RIS, reconfigurable intelligent surfaces, holographic MIMO, iterative detection and decoding, channel estimation.
\end{IEEEkeywords}

\section{Introduction}

With the advent of sixth-generation (6G) wireless networks, the demand for ultra-high data rates continues to escalate. Among the emerging technologies proposed to meet these requirements is Holographic Multiple-Input Multiple-Output (HMIMO). Practical realizations of HMIMO rely on reconfigurable metasurfaces operating in the analog domain. This approach overcomes the hardware constraints of conventional massive MIMO systems \cite{mmimo,wence} by utilizing dense arrays of sub-wavelength elements with minimal radio-frequency (RF) chains~\cite{9136592}.

To fully exploit the potential of metasurfaces as large antenna apertures, recent studies have explored multi-layer configurations to provide additional degrees of freedom for electromagnetic control. By appropriately tuning the response of each layer, these architectures can perform advanced wave-domain processing operations~\cite{10534211}. A pioneering architecture in this domain is the Stacked Intelligent Metasurface (SIM)~\cite{10158690, 10130641}, which employs diagonally structured transmissive RIS layers. Extending this concept, recent studies have introduced Beyond-Diagonal (BD) metasurfaces at every layer~\cite{11062671}, allowing for general linear operations and substantially enhancing the manipulation of the incident electromagnetic field.

Despite these promising capabilities, metasurface-assisted HMIMO systems face a substantial increase in the number of channel coefficients, as each meta-atom introduces an additional propagation path. To address this, several channel estimation methods have been proposed. For instance,~\cite{10892229} develops a minimum mean-square error (MMSE)-based estimator for SIM-HMIMO under Rician fading;~\cite{11168823} introduces a low-complexity tensor-based protocol for joint channel estimation and inter-layer calibration; and~\cite{10445164} proposes a framework enabling accurate CSI recovery via LS, MMSE, and subspace-based estimators.

Motivated by these challenges, this work presents a code-aided framework inspired by that  in~\cite{11143508} for metasurface-based HMIMO systems, considering both SIM and BD-SIM architectures. Generalizing the uplink LDPC strategy, we develop a channel estimation protocol where both pilot and parity bits contribute to the iterative refinement of the high-dimensional HMIMO channel. Furthermore, we derive closed-form expressions for the metasurface parameter design and incorporate them into an alternating-optimization (AO) procedure. Numerical results demonstrate substantial gains in normalized mean square error (NMSE) and bit error rate (BER), with major improvements observed in BD-SIM architectures.

\textit{Notation:} Bold capital letters represent matrices, while bold lowercase letters denote vectors. The symbol $\mathbf{I_n}$ refers to an $n \times n$ identity matrix and $\text{diag}(\mathbf{A})$ is a diagonal matrix with the diagonal elements of $\mathbf{A}$. The sets of complex and real numbers are denoted by $\mathbb{C}$ and $\mathbb{R}$, respectively. [·]$^{-1}$, [·]$^{T}$, and [·]$^H$ denote the inverse, transpose, and conjugate transpose, respectively. The Kronecker product is represented by $\otimes$.

\begin{figure*}[h]
    \vspace{-1em}
    \centering
    \includegraphics[width=0.8\textwidth]{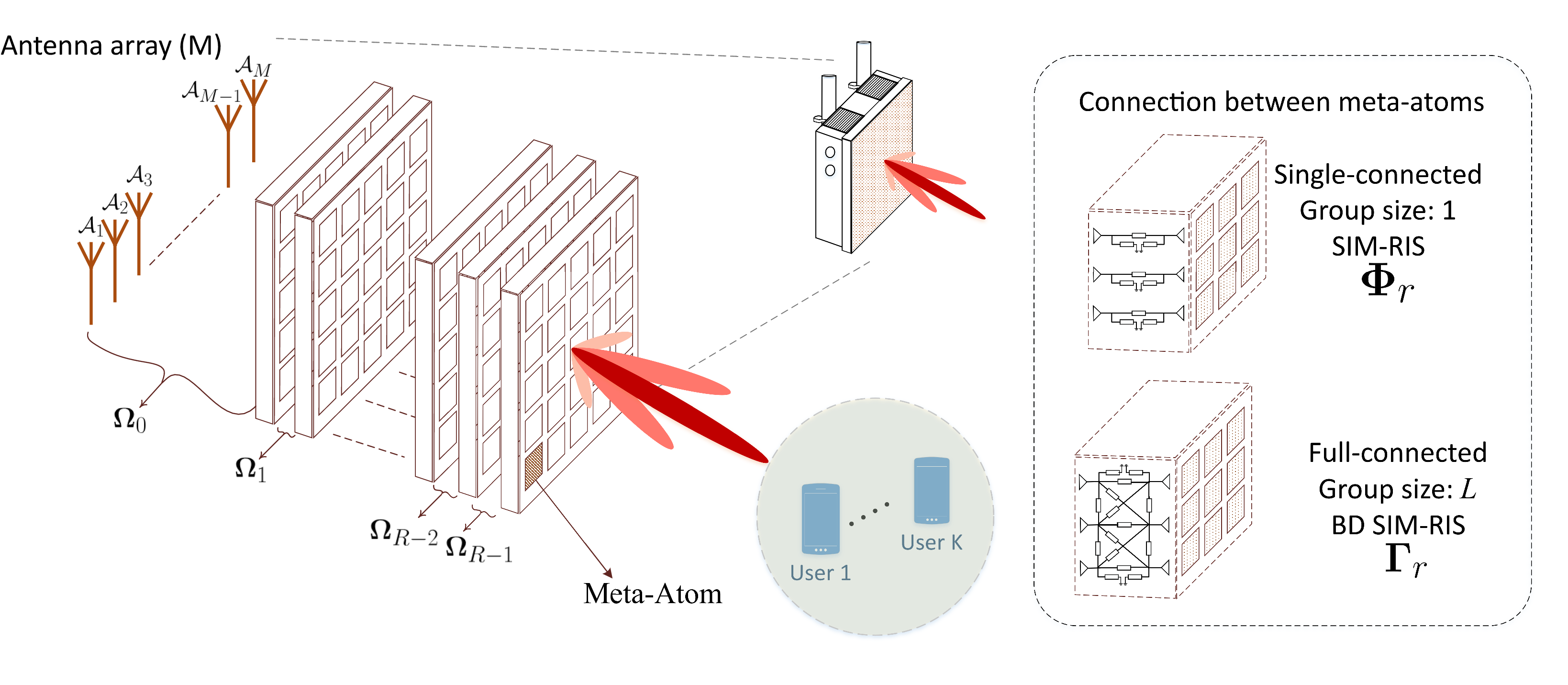}
    \vspace{-1.0em}
    \caption{System deployment illustrating the proposed HMIMO schemes.}
    \label{fig:deployment}
    \vspace{-0.5em}
\end{figure*}

\section{System Model}

\begin{figure}
    \vspace{-1em}
    \includegraphics[width=0.5\textwidth]{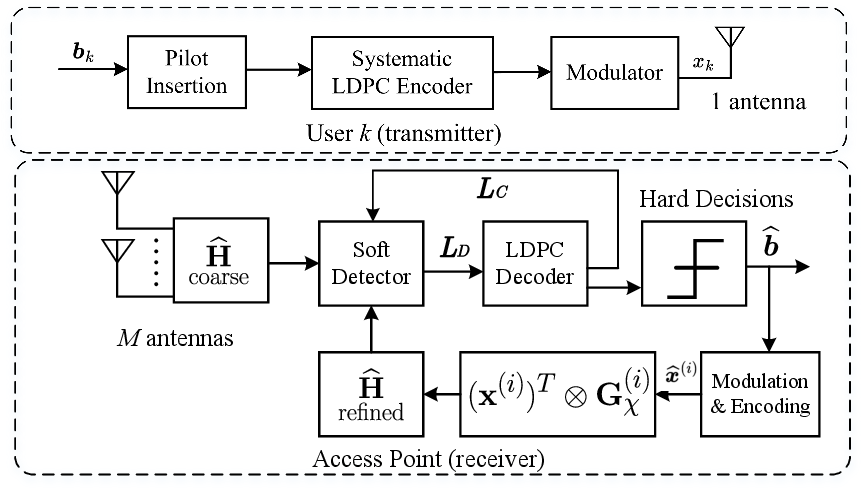}
    \vspace{-1.0em}
    \caption{Block diagram of the uplink scheme.}
    \label{fig:block_diagram}
    \vspace{-0.5em}
\end{figure}

The information bits of each user are encoded using individual LDPC encoders and subsequently modulated into symbols $x_k$ via QPSK modulation, as illustrated in Fig.~\ref{fig:block_diagram}. 
The transmitted symbols $x_k$ are zero-mean and have identical power, satisfying $E[|x_k|^2] = \sigma_x^2$. 
Transmission occurs over block-fading channels.

The $M$-dimensional received signal at the AP at time instant $i$ is expressed as
\begin{equation}
    \mathbf{y}^{(i)} = \mathbf{G}_\chi^{(i)} \mathbf{H} \mathbf{x}^{(i)} + \mathbf{n}^{(i)},
    \label{eq01}
\end{equation}
where 
$\mathbf{H} \in \mathbb{C}^{N \times K}$ represents the uplink channels from the $K$ users to the outermost metasurface; 
$\mathbf{G}_{\chi}^{(i)} \in \mathbb{C}^{M \times N}$ denotes the wave transformation matrix between the metasurface and the AP for architecture 
$\chi \in \{\text{SIM}, \text{BD}\}$; 
$\mathbf{x}^{(i)} \triangleq [x_1^{(i)}, \dots, x_K^{(i)}]^{\mathrm{T}}$ is the transmitted symbol vector; and $\mathbf{n}^{(i)} \sim \mathcal{CN}(\mathbf{0}_M, \sigma_n^2\mathbf{I}_M)$ represents the additive white Gaussian noise (AWGN). 
The superscript $(i)$ indicates the discrete transmission block index.

In both architectures, the RIS consists of $R$ stacked metasurface layers, each comprising $L$ reconfigurable elements capable of dynamically tuning their electromagnetic response under the control of an intelligent controller, thereby enabling wave-based spatial processing~\cite{10922857,11062671,iced}. 

The following subsections detail each architecture.

\subsection{Stacked Intelligent Metasurface Model}

In the SIM architecture, each layer is modeled as a complex coefficient vector
\begin{equation}
\boldsymbol{\varphi}_{r}^{(i)} = [e^{j\theta_{r,1}^{(i)}}, \dots, e^{j\theta_{r,L}^{(i)}}]^{\mathrm{T}},
\end{equation}
where $\theta_{r,\ell}^{(i)}$ denotes the phase shift of the $\ell$-th meta-atom on the $r$-th layer. Each layer can also be represented by a diagonal phase matrix
\begin{equation}
    \mathbf{\Phi}_{r}^{(i)} = \mathrm{diag}(\boldsymbol{\varphi}_{r}^{(i)}),
\end{equation}

The propagation between adjacent metasurface surfaces is represented by a transmission matrix $\boldsymbol{\Omega}_r \in \mathbb{C}^{L \times L}$, which characterizes the link between the $r$-th and $(r{+}1)$-th metasurfaces, as illustrated in Fig.~\ref{fig:block_diagram}. 
For the connection between the antenna array and the first surface ($r = 0$), the corresponding transmission matrix is $\mathbf{\Omega}_0 \in \mathbb{C}^{M \times L}$ ~\cite{10158690}. 
Each entry $\omega_{l,l'}^{(r)}$ models the coupling between the $l$-th element of layer $r$ and the $l'$-th element of layer $r+1$ and is given by
\begin{equation}
\omega_{l,l'}^{(r)} = 
\frac{A \cos \Theta_{l,l'}^{(r)}}{d_{l,l'}^{(r)}}
\left(\frac{1}{2\pi d_{l,l'}^{(r)}} - j\frac{1}{\lambda}\right)
e^{j 2\pi d_{l,l'}^{(r)}/\lambda},
\label{eq:t_matrix}
\end{equation}
where $\lambda$ is the wavelength, $A$ is the area of each meta-atom, 
$d_{l,l'}^{(r)}$ is the transmission distance, and $\Theta_{l,l'}^{(r)}$ is the angle between the propagation path and the normal to the metasurfaces. 

Therefore, the SIM wave transformation matrix $\mathbf{G}^{(i)}_\text{SIM} \in \mathbb{C}^{M \times N}$ is given by
\begin{equation}
    \label{eq:G_sim}
    \mathbf{G}^{(i)}_\text{SIM} =
    \mathbf{\Omega}_{0}\mathbf{\Phi}_{1}^{(i)} \mathbf{\Omega}_{1} 
    \mathbf{\Phi}_{2}^{(i)} \cdots 
    \mathbf{\Omega}_{R-1} \mathbf{\Phi}_{R}^{(i)}.
\end{equation}

\subsection{Beyond-Diagonal SIM Model}

In a BD-RIS, each element of the surface is interconnected with all others through a reconfigurable impedance network \cite{9913356,10308579,10587164}. 
We assume that all elements are mutually connected, so that the coefficient matrix of each layer $r$ can be represented as $\mathbf{\Gamma}_r \in \mathbb{C}^{L \times L}$, 
satisfying $\mathbf{\Gamma}_r\mathbf{\Gamma}_r^H = \mathbf{I}_L$~\cite{9913356}. 
For the propagation between the active antenna array and the RIS, as well as between consecutive metasurface layers, the same transmission-matrix model used for the SIM-RIS case is adopted.

Consequently, the BD wave transformation matrix $\mathbf{G}^{(i)}_\mathrm{BD} \in \mathbb{C}^{M \times L}$ is expressed as
\begin{equation}
    \label{eq:G_BD}
    \mathbf{G}^{(i)}_\text{BD} =
    \mathbf{\Omega}_{0}\mathbf{\Gamma}^{(i)}_1 \mathbf{\Omega}_{1} 
    \mathbf{\Gamma}^{(i)}_2 \cdots 
    \mathbf{\Omega}_{R-1} \mathbf{\Gamma}^{(i)}_R.
\end{equation}

\subsection{SIC-Based Linear Detection}
\label{subsec:SIC}

At the receiver, an estimate $\hat{x}_k$ of the $k$th transmitted symbol is obtained by applying a linear receive filter $\mathbf{w}_k$ to the received signal $\mathbf{y}^{(i)}$. 
Without performing soft interference cancellation (SIC), the conventional linear detector output is expressed as
\begin{equation}
    \hat{x}_k^{(i)} = (\mathbf{w}_k^{(i)})^H\mathbf{y}^{(i)} 
    = \left(\frac{\sigma^2_n}{\sigma^2_x}\mathbf{I}_{n_r} 
    + \mathbf{\bar{H}}_\chi \mathbf{\bar{H}}_\chi^{\rm H} \right)^{-1}
    \mathbf{\bar{h}}_{\chi,k}\mathbf{y}^{(i)},
    \label{detection_estimate_1}
\end{equation}
where 
\begin{equation}
    \mathbf{\bar{H}}_\chi = \mathbf{G}_\chi \mathbf{H}
\end{equation} represents the equivalent uplink channel between the AP and the users, 
$\mathbf{\bar{h}}_{\chi,k} \in \mathbb{C}^{M}$ is the $k$th user channel vector, and 
$\mathbf{\bar{H}}_\chi \triangleq [\mathbf{\bar{h}}_{\chi,1}, \dots, \mathbf{\bar{h}}_{\chi,K}] \in \mathbb{C}^{M \times K}$.

Although this linear detector provides a straightforward solution, its performance degrades under strong multiuser interference. 
To mitigate this effect, we employ a SIC strategy \cite{jiols,jiomimo,mfsic,dfcc,detmtc,listmtc,msgamp,msgamp2,iddllr,decidd,iddocl,rsrbd,rsthp}, which iteratively refines the symbol estimates by subtracting the reconstructed interference from previously detected users.

In the SIC detector \cite{spa,mfsic,mbdf}, the received vector $\boldsymbol{y}^{(i)}$ is first processed by a demapper that computes the log-likelihood ratio (LLR) for each coded bit in $\boldsymbol{x}^{(i)}$. 
For simplicity, the time index $(i)$ is omitted in this subsection. 
The extrinsic LLR value $L_D$ for the $\upsilon$th code bit $b_\upsilon$ is computed as
\begin{equation}
    L_{D}(b_{\upsilon}) = 
    \log \frac{
        \sum\nolimits_{\boldsymbol{x} \in \mathcal{X}_{\upsilon}^{+1}} 
        P(\boldsymbol{y} \vert \boldsymbol{x}, \mathbf{\bar{H}}_\chi) P(\boldsymbol{x})
    }{
        \sum\nolimits_{\boldsymbol{x} \in \mathcal{X}_{\upsilon}^{-1}} 
        P(\boldsymbol{y} \vert \boldsymbol{x}, \mathbf{\bar{H}}_\chi) P(\boldsymbol{x})
    } 
    - L_{C}(b_{\upsilon}),
    \label{eq:ldlc}
\end{equation}
where $L_C(b_{\upsilon})$ is the a priori LLR provided by the channel decoder.

Following the principles of iterative detection and decoding (IDD) \cite{774855,spa,mfsic,mbdf,bfidd,idd1bit,rrmber,iced}, 
a soft estimate of the $k$th transmitted symbol is obtained from the extrinsic information $\boldsymbol{L}_C$ as
\begin{equation*} 
\tilde {x}_{k} = \sum _{x\in \mathcal {D}} x\, \text{Pr}(x_{k}=x)
= \sum _{x\in \mathcal {D}} x 
    \left(\prod _{l=1}^{M_{c}}\! \big[1+\text{exp}(-x^{l}L_{c}^{l})\big]^{-1}\right),
\end{equation*}
where $\mathcal{A}$ denotes the modulation constellation with $2^{M_c}$ symbols, 
and $x^l \in \{+1, -1\}$ represents the $l$th bit of symbol $x$.

In the SIC stage, each symbol estimate $\tilde{x}_k$ is refined by minimizing the mean square error (MSE) between the transmitted symbol and the filter output, assuming the current RIS configuration $\boldsymbol{\varphi}$ is fixed. 
The optimal linear receive filter $\mathbf{w}_k$ \cite{jidf} is thus obtained as
\begin{equation} 
    \mathbf{w}_k = 
    \arg \min _{ \tilde{\mathbf{w}}_k} 
    E\!\left [{\left | x_k - \tilde{\mathbf{w}}_k^{H}\mathbf{y}\right | ^{2}}\right],
\end{equation}
whose closed-form solution is given by
\begin{equation}
    \mathbf{w}_k = 
    \left(\frac{\sigma^2_n}{\sigma^2_x}\mathbf{I}_{n_r} 
    + \mathbf{\bar{H}}_\chi \mathbf{\Delta}_k \mathbf{\bar{H}}_\chi^{\rm H} \right)^{-1}
    \boldsymbol{\bar{h}}_k,
    \label{eq:w}
\end{equation}
where $\mathbf {\bar{H}}_p \triangleq [\mathbf{\bar{h}}_1, \dots, \mathbf{\bar{h}}_K]^H$ is the equivalent channel matrix, 
and $\mathbf{\Delta}_k$ is a covariance weighting matrix defined as
\begin{equation}
    \mathbf{\Delta}_k = 
    \mathrm{diag}\!\left[
        \frac{\sigma^2_{x_{1}}}{\sigma^2_x}, \dots, 
        \frac{\sigma^2_{x_{k-1}}}{\sigma^2_x}, 1, 
        \frac{\sigma^2_{x_{k+1}}}{\sigma^2_x}, \dots, 
        \frac{\sigma^2_{x_{K}}}{\sigma^2_x} 
    \right],
\end{equation}
where $\sigma^2_{x_{k}}$ is the variance of the $k$th user symbol, computed as
\begin{equation}
\sigma_{x_{k}}^{2} = 
\sum\nolimits_{x\in \mathcal{D}} 
|x - \bar{x}_{k}|^{2} P(x_{k}=x).
\end{equation}

\section{Proposed Design of RIS Parameters}
\label{sec:RIS_par}

This section describes the optimization of the reconfigurable surface parameters for holographic MIMO systems. The proposed design employs an AO framework to handle multilayer metasurfaces, and the parameter update of each layer is then specialized based on whether the metasurface is diagonal or beyond-diagonal. For simplicity, the time index $(i)$ is also omitted in section.

\subsection{General Multilayer Optimization Framework}

Consider an RIS consisting of $R$ transmissive layers. During each AO iteration, only one layer is optimized while the remaining layers are kept fixed. Let $r$ denote the layer being optimized. An estimate $\hat{x}_k$ of the transmitted symbol of user $k$ is obtained by applying a linear receive filter $\mathbf{w}_k$ to the received signal:
\begin{equation}
     \hat{x}_k
     =
     \mathbf{w}_k^\mathrm{H}
     \underbrace{\left(
        \mathbf{\Omega}_{0}
        \mathbf{T}_{1} \mathbf{\Omega}_{1}
        \mathbf{T}_{2} \cdots
        \mathbf{\Omega}_{R-1} \mathbf{T}_{R}
        \mathbf{H}
     \right)}_{\mathbf{A}_r \mathbf{T}_r \mathbf{C}_r}
     \mathbf{x}
     +
     \mathbf{w}_k^{\mathrm{H}}\mathbf{n},
     \label{eq:ml_estimate}
\end{equation}
where the effective matrices for layer $r$ are
\begin{equation}
    \mathbf{A}_r
    =
    \mathbf{\Omega}_{0}
    \left(
        \prod_{t=1}^{r}
        \mathbf{T}_{t}\mathbf{\Omega}_{t}
    \right),
\end{equation}
\begin{equation}
    \mathbf{C}_r
    =
    \left(
        \prod_{t=r}^{R-1}
        \mathbf{\Omega}_{t+1}\mathbf{T}_{t+1}
    \right)\mathbf{H}.
\end{equation}
Note that only the operator of the $r$th layer, denoted by $\mathbf{T}_r$, is updated, while all other layers remain fixed. The matrix $\mathbf{T}_r$ contains the reflection parameters of the $r$th layer, which may take a diagonal or full-matrix structure depending on the architecture. After computing the effective matrices $\mathbf{A}_r$ and $\mathbf{C}_r$, the update of $\mathbf{T}_r^{(i)}$ becomes a single-layer optimization problem. The update rule depends on whether the layer follows the SIM-RIS or BD-RIS model, as described next.

\subsection{Layer-Specific Parameter Update}

\subsubsection{SIM-RIS}
For meta-surfaces with diagonal reflection matrix, the operator becomes $\mathbf{T}_r = \mathbf{\Phi}_r = \mathrm{diag}(\boldsymbol{\varphi}_r)$, and the reflection vector is obtained following the MMSE design in~\cite{10747209}. Defining the receive filter matrix $\mathbf{W} = [\mathbf{w}_1,\dots,\mathbf{w}_K]^H \in \mathbb{C}^{K\times M}$, the MMSE solution is
\begin{equation}
    \boldsymbol{\varphi}_{r,\mathrm{opt}} = \mathbf{B}^{-1}\boldsymbol{\Psi},
\end{equation}
where
\begin{equation}
    \mathbf{B}
    =
    \sum_{k=1}^{K}
    \mathrm{diag}(\mathbf{c}_{r,k})^{H}
    \mathbf{A}_r^{H}\mathbf{W}^{H}\mathbf{W}\mathbf{A}_r
    \mathrm{diag}(\mathbf{c}_{r,k}),
\end{equation}
\begin{equation}
    \boldsymbol{\Psi}
    =
    \sum_{k=1}^{K}
    \mathrm{diag}(\mathbf{c}_{r,k})^{H}
    \mathbf{A}_r^{H}\mathbf{W}^{H}
    \bigl(\mathbf{e}_k - \mathbf{W}\bar{\mathbf{h}}_k\bigr),
\end{equation}
with $\mathbf{c}_{r,k}$ denoting the $k$th column of $\mathbf{C}_r$. Note that this solution is not constrained to the unit-modulus restriction. To satisfy the constraint, the coefficients are projected as
\begin{equation}
[\boldsymbol{\varphi}_{r,\mathrm{trunc}}]_l = \frac{[{\varphi}_{r,\text{opt}}]_l}{|[{\varphi}_{r,\text{opt}}]_l|}  \Leftrightarrow  \boldsymbol{\varphi}_{r,\mathrm{trunc}}= e^{j\measuredangle (\boldsymbol{\varphi_{r,\text{opt}}})}.
\end{equation}
This phase-projection provides the closest feasible approximation to the MMSE solution while preserving low complexity and near-optimal performance.

\subsubsection{Beyond-Diagonal RIS}

 The operator $\mathbf{T}_r$ becomes a full reflection matrix denoted by $\mathbf{\Gamma}_r$, and the received signal model in \eqref{eq:ml_estimate} yields

\begin{equation}
    \hat{\mathbf{x}}
    =
    \mathbf{W}^{H}
    \left(
        \mathbf{A}_r \mathbf{\Gamma}_r \mathbf{C}_r
    \right)\mathbf{x}
    +
    \mathbf{W}^{H}\mathbf{n}.
\end{equation}
Unlike SIM-RIS, where only the diagonal was optimized, here we focus on optimizing the whole matrix. Assuming orthogonal pilot transmission, the unconstrained least-squares estimate reduces to
\begin{equation}
    \mathbf{\Gamma}_\mathrm{o} = 
    \left(
        \mathbf{C}_r\mathbf{W}^\mathrm{H}\mathbf{A}_\mathrm{r}
    \right)^{+}.
\end{equation}

To satisfy the RIS constraint 
$\mathbf{\Gamma}\mathbf{\Gamma}^\mathrm{H} = \mathbf{I}_L$, 
the following optimization problem is solved:
\begin{align}
    \mathbf{\Gamma}_t 
    = 
    \underset{\mathbf{\Gamma}_t}{\mathrm{arg\,min}} \;
    \|\mathbf{\Gamma}_\mathrm{o} - \mathbf{\Gamma}_t\|_F^2,
    \quad 
    \text{s.t.} \quad
    \mathbf{\Gamma}_t\mathbf{\Gamma}_t^\mathrm{H} = \mathbf{I}_L.
\end{align}

Applying the singular value decomposition (SVD), 
$\mathbf{\Gamma}_\mathrm{o} 
= 
\mathbf{U}\boldsymbol{\Sigma}_{\Gamma_\mathrm{o}}\mathbf{V}^\mathrm{H}$, 
and assuming that $\mathbf{\Gamma}_t$ shares the same basis, the problem reduces to  
\begin{equation}
    \underset{\boldsymbol{\Sigma}_{\Gamma_t}}{\mathrm{min}} \;
    \|\boldsymbol{\Sigma}_{\Gamma_\mathrm{o}} - \boldsymbol{\Sigma}_{\Gamma_t}\|_F^2,
\end{equation}
subject to
\begin{equation}
    \mathbf{\Gamma}_t\mathbf{\Gamma}_t^\mathrm{H} 
    =
    \mathbf{U}\boldsymbol{\Sigma}_{\Gamma_t}
    \boldsymbol{\Sigma}_{\Gamma_t}^\mathrm{H}
    \mathbf{U}^\mathrm{H}
    = 
    \mathbf{I}_L.
\end{equation}

Since $\boldsymbol{\Sigma}_{\Gamma_t}$ must be real and diagonal, the only feasible solution is $\boldsymbol{\Sigma}_{\Gamma_t} = \mathbf{I}_L$, yielding  
\begin{equation}
    \mathbf{\Gamma}_t = \mathbf{U}\mathbf{V}^\mathrm{H}.
    \label{eq:bd_trunc}
\end{equation}

\section{Proposed Code-Aided Channel Estimation}
\label{sec:CA_CE}

\begin{figure}
    \centerline{\includegraphics[width=0.45\textwidth]{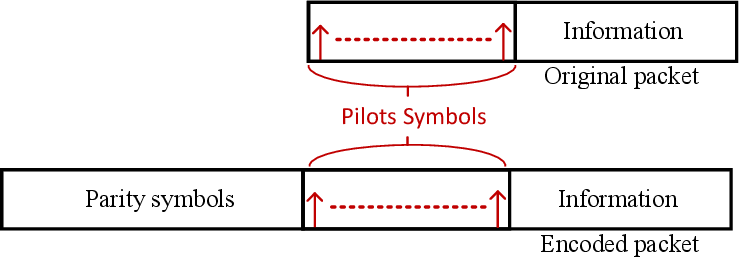}}
    \vspace{-0.725em}
    \caption{Systematically encoded packet and corresponding post-modulation structure.}
    \label{fig:ep}
    \vspace{-1em}
\end{figure}

This section presents the formulation of the proposed iterative code-aided channel estimation scheme for metasurface-based HMIMO systems. The method integrates IDD with encoded pilots (EP) and LDPC decoding to reduce pilot overhead while improving the accuracy of the cascaded RIS channel estimation. 
As illustrated in Fig.~\ref{fig:ep}, all pilot symbols are embedded within a systematic encoder together with the information bits, so that the pilot positions remain unaltered and are known at the receiver. 

The estimation procedure starts with an initial coarse estimate of the channel using a small number of pilot symbols. 
After decoding, the reconstructed packet is re-encoded and remodulated, allowing the estimated symbols to serve as new pilots for subsequent iterations. Since erroneous symbol estimates can propagate, the estimation is refined iteratively, updating both channel and data estimates until convergence or until a predefined number of iterations is reached.
To introduce this approach, we first describe the channel estimation for Holographic MIMO. Then, we show how the integration of encoded pilots and IDD enhances estimation accuracy. 

\subsection{Channel Estimation for Holographic MIMO}

Consider a receive symbol vector at the time instant $i$ expressed as
\begin{equation}
    \mathbf{y}^{(i)} 
    = \mathbf{G}_\chi^{(i)} \mathbf{H} \mathbf{x}^{(i)} + \mathbf{n}^{(i)},
    \label{eq:rx_mtx}
\end{equation}

To facilitate the derivation,~\eqref{eq:rx_mtx} is vectorized as
\begin{equation}
      \mathrm{vec}(\mathbf{y}^{(i)} )    
    = \left((\mathbf{x}^{(i)})^T \otimes \mathbf{G}_\chi^{(i)}\right) 
      \mathrm{vec}(\mathbf{H}) 
      + \mathrm{vec}(\mathbf{n}^{(i)}),
    \label{eq:rx_mtx_vec}
\end{equation}
Defining $(\boldsymbol{\lambda}^{(i)})^T = (\mathbf{x}^{(i)})^T \otimes \mathbf{G}^{(i)}_{\chi}$ and collecting samples at $\tau$ instants of time, we obtain
\begin{equation}
    \mathbf{\Lambda}_\tau =
    \begin{bmatrix}
        (\boldsymbol{\lambda}^{(i)})^T \\
        (\boldsymbol{\lambda}^{(i+1)})^T \\
        \vdots \\
        (\boldsymbol{\lambda}^{(i+\tau-1)})^T
    \end{bmatrix}
    \in \mathbb{C}^{\tau \times KN},
    \label{eq:lambda_matrix}
\end{equation}
such that~\eqref{eq:rx_mtx_vec} can be compactly rewritten as
\begin{equation}
    \mathrm{vec}(\mathbf{Y}_\tau) = 
    \mathbf{\Lambda}_\tau \, \mathrm{vec}(\mathbf{H}) + 
    \mathrm{vec}(\mathbf{N}_\tau).
    \label{eq:rx_mtx_vec_final}
\end{equation}
where $\mathbf{Y}_\tau = [\mathbf{y}^{(i)}, \dots, \mathbf{y}^{(i+\tau-1)}] \in \mathbb{C}^{M \times \tau}$ represent a transmission block composed of $\tau$ consecutive pilot symbol vectors, and $\mathbf{N}^{(i)} = [\mathbf{n}^{(i)}, \dots, \mathbf{n}^{(i+\tau-1)}] \in \mathbb{C}^{M \times \tau}$  the noise matrix.

Based on~\eqref{eq:rx_mtx_vec_final}, different channel estimators can be applied. In this work, we adopt the linear MMSE (LMMSE) estimator: 
\begin{equation}
    \mathrm{vec}(\hat{\mathbf{H}}) = 
    \mathbf{\Lambda}_\tau^H
\left(\mathbf{\Lambda}_\tau\mathbf{R}_\Lambda\mathbf{\Lambda}_\tau^H
    + \frac{\sigma_n^2}{\sigma_x^2}\mathbf{I}\right)^{-1}
    \mathrm{vec}(\mathbf{Y}_p),
    \label{eq:lmmse}
\end{equation}
where $\mathbf{R}_\Lambda = \mathbb{E}[\mathbf{\Lambda}_\tau\mathbf{\Lambda}_\tau^H] \in \mathbb{C}^{KN \times KN}$ 
denotes the channel covariance matrix.

To mitigate multiuser interference, $\mathbf{\Lambda}$ should ideally be orthogonal or semi-orthogonal, ensuring well-conditioned inversion and avoiding noise amplification. 
To achieve this, we generate $\boldsymbol{\lambda}$ using Hadamard sequences for $\mathbf{x}^{(i)}$, while $\mathbf{G}^{(i)}_{\chi}$ is obtained by assigning random values to the metasurface coefficients.
However, full orthogonality may be infeasible since $\mathbf{\Lambda}$ must have rank at least $KL$. 
Therefore, we set the training length to $\tau = p$ and employ a limited number of pilot symbols, with $p \ll KL$, which yields a coarse approximation of the cascaded RIS channel; this imperfect initial estimate is then exploited as prior information for an iterative refinement process.

\subsection{Iterative Estimation Process}
\label{sec:proposedInt}

In the proposed framework, the decoder output from the IDD receiver provides soft symbol estimates, which are exploited as virtual pilots in subsequent iterations. 
As shown in Fig.~\ref{fig:block_diagram}, the decoded bits are re-encoded and modulated to reconstruct $\hat{\mathbf{x}}^{(i)}$. 
We then compute the Kronecker product $((\mathbf{\hat{x}}^{(i)})^T \otimes \mathbf{G}_\chi^{(i)}) $ to obtain $\hat{\boldsymbol{\Lambda}}$, enabling the LMMSE re-estimation of the channel in~(\ref{eq:lmmse}). 
It is crucial that $\hat{\boldsymbol{\Lambda}}$ remains semi-orthogonal and well-conditioned; therefore, a suboptimal design is adopted for the parity symbols to maintain pseudo-orthogonality. Therefore, we adopt a hybrid configuration strategy for the metasurfaces. 

\begin{algorithm}[H]
\caption{Encoded Pilot Packet Generation}
\label{alg:generation}
\begin{algorithmic}[1]
    \REQUIRE Input bitstream $\mathbf{b}_{k}$ for each user $k$
    \ENSURE Symbol stream for user $k$: $[x_k^{(1)}, \dots, x_k^{(N_{\text{extra}}+ N_{\text{info}})}]$
    
    \FOR{$k = 1$ to $K$}
        \STATE Generate auxiliary pilot symbols for user $k$: 
        \textcolor{black}{$\mathbf{x}_{\text{aux}} = S_\mathcal{A} \cdot \text{Hadamard}(k)$}
        
        \STATE Generate pilot bits: \textcolor{black}{$\mathbf{b}_p = \text{demodulate}(\mathbf{x}_{\text{aux}})$}
        
        \STATE Assemble pilot+data bitstream: $\mathbf{b}_{\text{packet},k} = [\mathbf{b}_p \: ; \: \mathbf{b}_{k}]$
        
        \STATE Encode: $\mathbf{c}_k = \text{encode}(\mathbf{b}_{\text{packet},k})$
        
        \STATE Modulate: $[x_k^{(1)}, \dots, x_k^{(N_{\text{extra}} + N_{\text{info}})}] = \text{modulate}(\mathbf{c}_k^\top)$
    \ENDFOR
\end{algorithmic}
\vspace{0.5em}
{\scriptsize 
\textbf{Observation:}
$N_{\text{extra}}$ denotes the length of the combined parity and pilot symbols, while 
$p$ and $N_{\text{info}}$ represent the pilot and data symbol lengths, respectively. 
$S_{\mathcal{A}}$ denotes an arbitrary constellation symbol from $\mathcal{A}$, and 
$\mathrm{Hadamard}(k)$ refers to the $k$-th row of a Hadamard matrix. The resulting pilot vectors are mutually orthogonal across users, facilitating interference-free separation at the receiver.
}
\end{algorithm}

For the pilot and parity symbol durations, we employ time-varying random phase configurations at each metasurface layer. This random variation provides the necessary rank and diversity to effectively probe the high-dimensional channel, ensuring that $\hat{\boldsymbol{\Lambda}}$ acts as a robust sensing matrix. Conversely, for the information (data) symbols, the metasurfaces are fixed to the optimized coefficients derived in Section~\ref{sec:RIS_par}, aiming to maximize the spectral efficiency during payload transmission. Consequently, the sequence of matrices $\mathbf{G}_\chi$ encapsulates both the random sensing states (during pilot/parity slots) and the optimized communication state (during data slots).

Although the initial symbol estimates may include residual errors that could propagate to the channel estimate, the proposed scheme mitigates this effect through iteration. The channel is refined by reprocessing the decoded symbols until convergence or until a predefined maximum number of iterations is reached. The pseudocode of the proposed ICCE algorithm is summarized below.

\begin{algorithm}[H]
\caption{Channel Estimation and Refinement}
\footnotesize
\label{alg:generation}
\begin{algorithmic}[1]
    \REQUIRE $\boldsymbol{\Lambda}$, ${\boldsymbol{\Theta}}_{\mathrm{ps}}$, ${\boldsymbol{\Theta}}_{*}$, ${\boldsymbol{\Theta}}_{o}$
    \ENSURE Estimated channels: $\hat{\mathbf{H}}$
    
    \STATE \textbf{Coarse Estimation:}
    \textcolor{black}{
    \begin{itemize}        
        \item Form matrix: $\mathbf{Y}_{p} = \left[ \mathbf{y}^{(i)}, \dots, \mathbf{y}^{(i - 1 + p)} \right]$
        \item Coarse channel estimation:
        \vspace{-0.2cm}
        \begin{equation*}
            \mathrm{vec}(\hat{\mathbf{H}}) = 
            \mathbf{\Lambda}_{p}^H
            \left(\mathbf{\Lambda}_{p}\mathbf{R}_\Lambda\mathbf{\Lambda}_{p}^H
            + \frac{\sigma_n^2}{\sigma_x^2}\mathbf{I}\right)^{-1}
            \mathrm{vec}(\mathbf{Y}_{p}).
            \label{eq:lmmse}
        \end{equation*}
    \end{itemize}}
    
    \vspace{0.1cm}
    \STATE \textbf{Iterative Estimation Process:}
    \FOR{$t = 1$ to Max Iterations or until $\text{NMSE} < \text{tol}$}
        \vspace{0.1cm}
        \STATE \textbf{IDD Scheme (SIC):} Apply the iterative detection and decoding as described in Sec.~\ref{subsec:SIC}
        \vspace{0.1cm}
        \STATE \textbf{Refinement:}
        \begin{itemize}
            \item Build:
            \vspace{-0.1cm}
            \begin{equation*}
                \boldsymbol{\hat{\Lambda}} = {\left[ (\mathbf{\hat{x}}^{(1)})^T \otimes \mathbf{G}_\chi^{(1)}, \dots, (\mathbf{\hat{x}}^{(N_\text{extra})})^T \otimes \mathbf{G}_\chi^{(N_\text{extra})} \right]}
            \end{equation*}            
            \vspace{-0.2cm}
            \item Form matrix: $\mathbf{Y}_{\text{extra}} = \left[ \mathbf{y}^{(i)}, \dots, \mathbf{y}^{(i - 1 + N_\text{extra})} \right]$
            \item Refined channel estimation:
            \vspace{-0.1cm}
        \begin{equation*}
            \mathrm{vec}(\hat{\mathbf{H}}_\mathrm{refined}) = 
            \mathbf{\hat{\Lambda}}^H
            \left(\mathbf{\hat{\Lambda}}\mathbf{R}_\Lambda\mathbf{\hat{\Lambda}}^H
            + \frac{\sigma_n^2}{\sigma_x^2}\mathbf{I}\right)^{-1}
            \mathrm{vec}(\mathbf{Y}_{\text{extra}}).
            \label{eq:lmmse}
        \end{equation*}
        \end{itemize}
        \vspace{-0.2cm}
    \ENDFOR
\end{algorithmic}
{\scriptsize 
\textbf{Observation:}
$N_{\text{extra}}$ denotes the length of the combined parity and pilot symbols and
$p$  represent the pilot symbol lengths. }
\end{algorithm}

\section{Numerical Results}

A short-length regular LDPC code \cite{bfpeg,memd} with a block length of $N=1024$ and rate $R = 1/2$ is considered in conjunction with QPSK modulation. 
The system operates under block fading assumptions, where the CSI is estimated at the receiver  during the uplink phase. 
To capture the electromagnetic coupling inherent to the dense deployment of meta-atoms in the SIM architecture, we adopt a spatially correlated Rician fading model for the channel matrix. The large-scale path loss follows the 3GPP TR 38.901 Urban Micro (UMi) - Street Canyon model for LOS propagation \cite{3gpp38901_2025}, given in dB by:
\begin{equation}
    PL(d) = 32.4 + 21 \log_{10}(d) + 20 \log_{10}(f_c) + \xi,
\end{equation}
where $d$ is the distance in meters, $f_c$ is the carrier frequency in GHz, and $\xi \sim \mathcal{N}(0, \sigma^2)$ denotes log-normal shadowing with $\sigma = 4$ dB. 
The remaining simulation parameters are summarized in Table \ref{tab:sim_params}.

\begin{table}[ht]
\centering
\caption{Simulation Parameters}
\label{tab:sim_params}
\begin{tabular}{l|c}
\hline
\textbf{Parameter} & \textbf{Value} \\ \hline
Carrier Frequency ($f_c$) & $6$ GHz \\ 
Setup & Uplink MU-MIMO \\ 
Scenario & Urban Micro (UMi) \\
RF Chains ($M$) & $8$ \\ 
RIS Elements ($N$) & $64$ ($8 \times 8$) \\ 
Element Spacing & $\lambda / 2$ \\ 
SIM Layers ($R$) & $1$ -- $5$ \\ 
Users ($K$) & $4$ \\ 
User Distance & $250$ m \\ 
Path Loss Model & $48.0 + 21\log_{10}(d)$ (dB) \\ 
User Distribution & Cluster (Radius: $10$ m) \\ 
Number of encoded Pilots & 32 \\ \hline
\textbf{Channel Model} & \textbf{Correlated Rician} \\ 
Rician Factor ($K_{Rice}$) & $3$ dB \\ 
Rx Correlation ($\mathbf{R}_{RX}$) & Sinc model \\ 
Tx Correlation ($\mathbf{R}_{TX}$) & Identity ($\mathbf{I}_K$) \\ \hline
\end{tabular}
\end{table}

\begin{figure}
\includegraphics[width=0.5\textwidth]{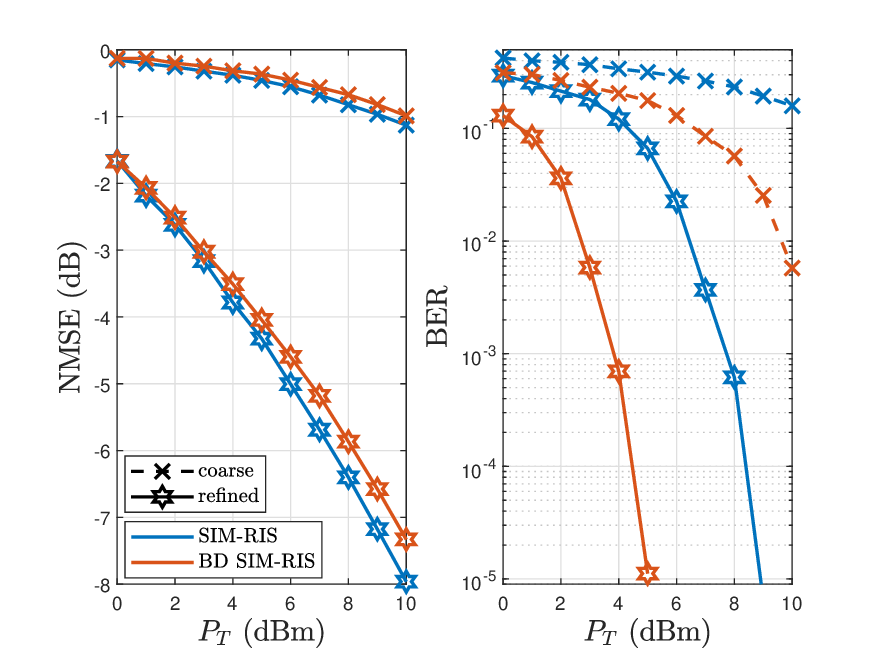}
    \caption{NMSE/BER Evaluation of Five-Layer SIM and BD-SIM system.}
    \label{fig:vtc_00}
\end{figure}

In the curves presented in Fig.~\ref{fig:vtc_00}, we compare the NMSE and BER performance of the BD-SIM and conventional SIM-RIS architectures, assuming five stacked layers for both designs. Here, $P_T$ denotes the transmit power per user. The BD-SIM configuration achieves slightly better performance as a result of the additional degrees of freedom enabled by its fully connected meta-atom structure. Furthermore, the figure displays the results for the coarse estimation followed by the refined performance obtained with a single iteration.

\begin{figure}
\includegraphics[width=0.5\textwidth]{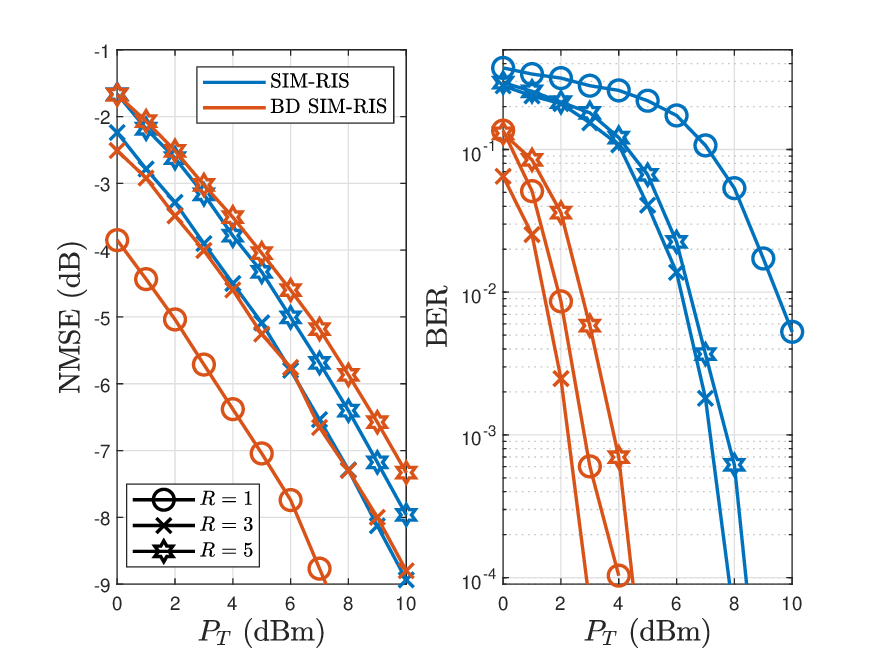}
    \caption{Influence of Layer Stacking on SIM-RIS/BD-SIM performance.}
    \label{fig:vtc_01}
\end{figure}

Fig.~\ref{fig:vtc_01} further evaluates the performance of both systems for different numbers of layers. When a single layer is considered ($R = 1$), the models reduce to the conventional BD-RIS and the conventional transmissive RIS cases. A general trend observed in both systems is the degradation of estimation accuracy as the number of layers increases. This behavior is attributed to the signal attenuation resulting from the sequential multiplication by the inter-layer terms $\mathbf{\Omega}_r$. Specifically regarding the BD-SIM, a performance decline is noted when increasing from three to five layers. This result suggests that the additional wave-domain directivity gain provided by the extra layers fails to offset the increased channel estimation error.

\section{Conclusion}

In this study, we proposed a novel iterative detection, decoding and channel estimation scheme for metasurface-based holographic MIMO systems. Unlike existing approaches, the proposed approach exploits uplink channel coding to leverage parity bits for both decoding and channel estimation, while employing encoded pilots to further enhance estimation accuracy. The scheme substantially reduces the minimum number of pilots required for reliable operation. We also derived a closed-form expression for metasurface parameter design using alternating optimization even though the proposed framework is not restricted to any specific optimization method. Numerical results demonstrate significant gains in channel estimation and BER. 

\newpage
    
\bibliography{support/main}
\end{document}